\documentclass{article}
\usepackage[a4paper, left=1.5cm, right=1.5cm, top=2.5cm, bottom=2.5cm]{geometry}
\usepackage{graphicx} % Required for inserting images
\usepackage{amsmath}
\usepackage{amssymb}
\usepackage{xcolor}
\usepackage{authblk}
\usepackage{cite}
\title{Future Dark Energy Constraints from Atomic Clocks}
\author{Oem Trivedi\thanks{oem.trivedi@vanderbilt.edu}}
\affil[]{Department of Physics and Astronomy, Vanderbilt University, Nashville, TN 37235, USA}

\date{\today}

\begin{document}

\maketitle

\begin{abstract}
We show that atomic clock measurements provides an exceptionally sensitive Solar System probe of scalar–tensor dark energy. By connecting variations in Newton’s constant and differential clock drifts to the dynamics of a single dark energy scalar, we derive a direct constraint on the present day equation of state and our results force any locally coupled scalar dark energy into a very slowly rolling regime with pushing $1+w_0\lesssim 10^{-5}$. This is independent of potential shape or kinetic structure and rules out broad classes of canonical and non-canonical models, leaving only near-$\Lambda$ behavior or fully decoupled fields as viable late-time scalar dark energy, thereby leaving cosmological constant and minimally coupled scalar field models as the most consistent dark energy regimes. We also use results from Lunar Laser Ranging and photon trajectories to further strengthen our the depth of our constraints.
\end{abstract}

\section{Introduction}
It would not be an exaggeration to say that modern cosmology is experiencing its most vibrant and transformative era. Across theory and observation, we are seeing unprecedented advances and those have enabled us to consider the Universe over immense dynamical ranges with remarkable precision. The development of Dark Energy models \cite{de1SupernovaSearchTeam:1998fmf,de2Li:2012dt,de3Li:2011sd,de4Mortonson:2013zfa,de5Frusciante:2019xia,de6Huterer:2017buf,de7Vagnozzi:2021quy,de8Adil:2023ara,de9Feleppa:2025clx,de10DiValentino:2020evt,de11Nojiri:2010wj,de12Nojiri:2006ri,de13Trivedi:2023zlf,de14Trivedi:2022svr,de15Trivedi:2024inb,de16Trivedi:2024dju}, for example, has drastically expanded the possible landscape of cosmological dynamics even as long standing anomalies such as the $H_0$ and $S_8$ tensions \cite{ht1DiValentino:2021izs,ht2Clifton:2024mdy,s81kazantzidis2018evolution,s82amon2022non,s83poulin2023sigma,s84Ferreira:2025lrd,baoFerreira:2025lrd} continue to challenge the standard $\Lambda$CDM model at a conceptual level. Simultaneously, we seeing an era of data driven cosmology is emerging, with next-generation CMB, LSS and high redshift probes, gravitational wave observatories and precision mapping of the cosmic web promising increasingly stringent tests of early Universe physics \cite{t1jha2019next,t2Shanks:2015lda,t3chandler2025nsf,t4Euclid:2024yrr,t5WST:2024rai,t6CosmoVerseNetwork:2025alb,t7COMPACT:2022gbl}. This correspondence between theoretical ideas and observational capability is perhaps building a new age of upheaval in the field as we ask fundamental questions about the nature of spacetime, gravity and cosmic evolution.
\\
\\
As theoretical possibilities for dark energy have proliferated, scalar field constructions have emerged as one of the most studied and versatile frameworks. The parameter space of dark energy models has become quite vast, ranging from canonical quintessence to non-minimal scalar–tensor theories, k-essence, Horndeski models, modified gravity inspired potentials and non-canonical kinetic sectors \cite{de2Li:2012dt,de3Li:2011sd}. Yet cosmology is not only shaped by large-scale surveys but at the same time, it is also strongly helped by Solar System and near-Earth tests which have long imposed powerful constraints on gravity, particle physics and dark sector interactions. For example, dark matter models are shaped by direct detection experiments and planetary ephemerides \cite{dm11rubin1970rotation,dm1Cirelli:2024ssz,dm2Arbey:2021gdg,dm3Balazs:2024uyj,dm4Eberhardt:2025caq,dm5Bozorgnia:2024pwk,dm6Misiaszek:2023sxe,dm7OHare:2024nmr,dm8Adhikari:2022sbh,dm9Miller:2025yyx,dm10Trivedi:2025vry} while pulsar timing arrays constrain ultra light fields and compact objects and precision orbital measurements delimit possible fifth forces \cite{pta1Gan:2025icr,pta2Unal:2022ooa,pta3Ramani:2020hdo}. Though such probes are rarely applied to dark energy given its ultra low density and apparent cosmological nature, one should still recognize that many theoretically motivated scalar field dark energy models predict small but non-zero effects on local gravitational dynamics or on the evolution of fundamental constants. This raises a compelling question which is as follows: Can Solar System measurements provide an independent and complementary window into dark energy physics?
\\
\\
The goal of this work is to explore precisely this possibility by combining two of the most sensitive local probes of gravitational and scalar-field dynamics, which are Lunar Laser Ranging (LLR) and Atomic Clock comparisons. LLR has previously been used to constrain dark energy motivated variations in Newton’s constant in scalar–tensor theories, while atomic clocks probe differential variations in particle masses and gauge couplings. Here we develop novel atomic clock constraints on scalar dark energy, obtaining what appears to be the strongest Solar System based constraint on the present day dark energy equation of state, and use constraints from LLR in conjunction too. In Section 2 we summarise the relevant aspects of LLR, PPN orbit constraints and derive their connections to the running of the effective Planck mass. Section 3 introduces atomic clock probes and establishes a novel relation connecting clock drifts and the dark energy equation of state. Sections 4 and 5 apply these results to canonical and non-canonical dark energy models respectively, giving us stringent constraints on scalar–tensor dark energy theories. We conclude in Section 6 with a discussion of the broader implications of our findings and potential directions for future work.
\\
\\
\section{Lunar Laser Ranging and PPN constraints on Planck Mass}
Lunar Laser Ranging is interesting as it provides one of the most precise long-baseline tests of gravitational physics in the Solar System \cite{llr1Williams:2004qba,llr2Williams:2004uw,llr3Merkowitz:2010kka,llr4Muller:2007zzb,llr5courde2017lunar}. It works by sending laser pulses from Earth then reflecting them off retroreflectors placed on the lunar surface and measuring the round trip travel time over several decades and it is possible to determine the Earth-Moon distance with millimeter accuracy. From this, one can infer the long term evolution of the lunar orbit and obtains stringent constraints on any possible temporal variation of the gravitational coupling. These constraints when interpreted within a scalar-tensor description of dark energy, can lead to quite strong restrictions on the allowed evolution of the effective Planck mass and therefore on the possible dynamics of dark energy at $z \simeq 0$.
\\
\\
We begin with a Jordan frame scalar-tensor action of the form
\begin{equation} \label{action}
S = \int d^4x\,\sqrt{-g}\left[\frac{1}{2}F(\phi)R - \frac{1}{2}(\nabla\phi)^2 - V(\phi)\right]
+ S_m[g_{\mu\nu},\Psi_m]
\end{equation}
where $F(\phi)$ encodes the nonminimal coupling between the scalar field and gravity and the quantity $M_*^2(\phi)=F(\phi)$ plays the role of an effective squared Planck mass while the effective Newton constant is given by
\begin{equation}
G_{\rm eff}(\phi) \simeq \frac{1}{8\pi F(\phi)}
\end{equation}
A useful parameter in the effective field theory description of dark energy is the running of the Planck mass
\begin{equation}
\alpha_M \equiv \frac{d\ln M_*^2}{d\ln a} = \frac{1}{H}\frac{\dot F}{F}
\end{equation}
Since $G_{\rm eff}\propto 1/F$, we see that any time variation of $F$ leads directly to \cite{llr9Tsujikawa:2019pih}
\begin{equation}
\frac{\dot G}{G} = -\frac{\dot F}{F} = -\alpha_M H
\label{eq:Gdot_basic}
\end{equation}
One should note what is interesting here is that this expression connects a local observable, the fractional rate of change of the gravitational coupling, to a cosmological quantity which is the running of the Planck mass. Now, on a homogeneous FLRW background we see that the scalar field contributes a dark energy density and pressure given by
\begin{equation}
\rho_\phi = \frac{1}{2}\dot\phi^2 + V(\phi) - 3H\dot F, \qquad
p_\phi = \frac{1}{2}\dot\phi^2 - V(\phi) + \ddot F + 2H\dot F
\end{equation}
If we now define $w_{\rm DE}=p_\phi/\rho_\phi$ and considering the regime in which the $F$ dependent contributions to the background are subdominant, we may therefore approximate
\begin{equation}
\rho_\phi \approx \frac{1}{2}\dot\phi^2 + V(\phi), \qquad
p_\phi \approx \frac{1}{2}\dot\phi^2 - V(\phi)
\end{equation}
We can then evaluate this today and get
\begin{equation}
1+w_0 = \frac{\dot\phi_0^2}{\rho_{\phi,0}}
\approx \frac{\dot\phi_0^2}{3M_{\rm Pl}^2H_0^2\Omega_{\rm DE,0}}
\end{equation}
or equivalently
\begin{equation}
\dot\phi_0^2 \approx 3\,\Omega_{\rm DE,0}\,(1+w_0)\,M_{\rm Pl}^2H_0^2
\label{eq:phidot_background}
\end{equation}
where we note that $\Omega_{\rm DE,0}$ is the energy density parameter for dark energy at $z=0$. Equations \eqref{eq:Gdot_basic} and \eqref{eq:phidot_background} will be central in the subsequent analysis as well as they link the background dynamics of dark energy to the local physics probed by LLR.
\\
\\
To make contact with LLR observations, we now consider the long term evolution of the lunar orbit where in Newtonian approximation, the orbital dynamics are governed by the simple Keplerian form
\begin{equation}
n^2 a^3 = G M
\end{equation}
where $n=2\pi/P$ is the mean motion, $a$ is the semi-major axis and $M$ is the total mass of the Earth-Moon system. If the gravitational coupling varies slowly with time $G(t)=G_0+\dot G\,t+\cdots$, then the orbit evolves adiabatically and so differentiating Kepler's relation and using approximate angular momentum conservation gives us
\begin{equation}
\frac{\dot a}{a} \simeq -\frac{\dot G}{G}, \qquad \frac{\dot n}{n} \simeq 2\frac{\dot G}{G}
\label{eq:orbital_drift}
\end{equation}
These relations imply that even a tiny fractional drift in $G$ gives us secular changes in the orbital elements and so the most direct observable is the mean anomaly which is given by 
\begin{equation}
M(t) = \int n(t)\,dt \simeq n_0 t + \frac{1}{2}\dot n\,t^2 + M_0
\end{equation}
this acquires a quadratic in-time correction proportional to $\dot G/G$ and this is precisely the type of deviation measured by LLR fits to lunar ranging data.
\\
\\
Current LLR measurements give a constraint of the form \cite{llr6alBurrage:2020jkj,llr7alMould:2014iga,llr8Zheng:2025vyv}
\begin{equation}
\left(\frac{\dot G}{G}\right)_0 = (2\pm 7)\times 10^{-13}\,{\rm yr}^{-1}
\end{equation}
this gives us the conservative bound
\begin{equation}
G_1 \equiv \left|\frac{\dot G}{G}\right|_0 \lesssim 7\times10^{-13}\,{\rm yr}^{-1}
\simeq 2.2\times10^{-20}\,{\rm s}^{-1}
\label{eq:Gdot_bound}
\end{equation}
Using the relation $\dot G/G=-\alpha_M H$ evaluated today gives
\begin{equation}
\left(\frac{\dot G}{G}\right)_0 = -\alpha_{M,0}H_0
\end{equation}
we note here that $\alpha_{M,0}$ is the value of $\alpha_{M}$ at $z=0$ and therefore
\begin{equation}
G_1 \approx |\alpha_{M,0}|\,H_0
\end{equation}
Assuming the Hubble constant lies in the observationally motivated interval $H_0\in[67,73]\,{\rm km\,s^{-1}\,Mpc^{-1}}$ \cite{Planck:2018vyg,shoesRiess:2021jrx}, corresponding to $H_0\in[2.17,2.37]\times10^{-18}\,{\rm s}^{-1}$, one immediately finds
\begin{equation}
|\alpha_{M,0}| \lesssim \frac{G_1}{H_0} \sim (0.94\text{-}1.0)\times10^{-2}
\end{equation}
This shows that the present-day running of the Planck mass is bounded to be at most of order a few times $10^{-3}$ to $10^{-2}$ per Hubble time and this constraint is extremely restrictive. Note that this also holds independently of any assumptions regarding atomic clock measurements that we will do later on. Also we would like to make it clear that LLR and atomic clock relations are independent of each other here.
\\
\\
In addition to the dynamical constraints obtained from massive bodies such as the Earth-Moon system, complementary bounds on time-varying gravitational couplings and scalar-tensor interactions can be inferred from photon trajectories in the Solar System as well. It is so because the propagation of light is sensitive to the spacetime geometry through null geodesics and in scalar-tensor theories one sees that this geometry is modified both directly through the metric and indirectly through the scalar dependence of the effective Planck mass \cite{ppn1Will:2005va,ppn2Dzuba:2024pri,ppn3Dent:2008ev,ppn4Hou:2017cjy,ppn5Zhang:2023nil}. In the Jordan frame, photons follow null geodesics of $g_{\mu\nu}$ and deviations from general relativity can manifest themselves through changes in post-Newtonian parameters and through potential temporal evolution of these parameters. In particular, one should note that light deflection, Shapiro time delay and frequency shifts provide clean probes of the gravitational sector that are largely insensitive to the internal structure of massive bodies \cite{ppn6Saito:2024xdx,ppn7Richarte:2025edh,ppn8Damour:1992we,ppn9Scharer:2014kya}.
\\
\\
At the level of the parametrized post-Newtonian expansion (PPN expansion), scalar-tensor theories predict deviations in the parameter $\gamma_{\rm PPN}$. This governs the spatial curvature produced by unit rest mass and directly controls photon trajectories and to see this, considering an action of the form \eqref{action}, one finds
\begin{equation}
\gamma_{\rm PPN}-1 = -\frac{F_{,\phi}^2}{F + 2F_{,\phi}^2}
\end{equation}
so we see that any nonminimal coupling $F(\phi)$ leads to modifications of light bending and time delay and high precision measurements of the Shapiro delay from spacecraft tracking experiments, most notably the Cassini mission, constrain $|\gamma_{\rm PPN}-1|\lesssim 2.3\times10^{-5}$ \cite{ppn10Bondarescu:2014vta}. This in turn ends up restricting the present day scalar coupling strength and indirectly bounds the allowed variation of $F(\phi)$ and when this is combined with the relation $\dot G/G=-\dot F/F$, these constraints limit not only the instantaneous deviation from general relativity but also the rate at which photon trajectories can evolve over time.
\\
\\
Moreover one can see that long term timing of photon signals exchanged between Earth, spacecraft and planetary targets is sensitive to secular drifts induced by a time-varying gravitational coupling. The round trip light travel time $\Delta t$ acquires corrections schematically of the form
\begin{equation}
\Delta t \simeq \Delta t_{\rm GR}\left[1 + \mathcal{O}(\gamma_{\rm PPN}-1)\right] + \mathcal{O}\!\left(\frac{\dot G}{G} t\right)
\end{equation}
so that even a tiny $\dot G/G$ can lead to cumulative effects over multi-decade baselines, analogous to those probed by LLR but involving null rather than timelike trajectories. Future laser ranging to planetary landers, interplanetary transponders and space based interferometric missions will also hence provide us with increasingly precise photon based tests of scalar-tensor gravity, which will end up offering an independent and complementary avenue to constrain $\alpha_{M,0}$ and the late time dynamics of dark energy alongside Lunar Laser Ranging. We end here by noting that the bounds derived from photon trajectory observables such as light deflection and Shapiro time delay discussed above are fully consistent with the LLR constraints on $\alpha_{M,0}$, as both classes of measurements probe the same underlying scalar mediated modifications of the effective Planck mass \cite{ppn11Anton:2025zer}. While LLR constrains the time variation of $G$ through the dynamics of massive bodies, photon based tests restrict the instantaneous and secular evolution of the spacetime geometry, jointly enforcing that $|\alpha_{M,0}|$ must remain at most of order $10^{-2}$ at the present epoch.

\section{Atomic Clocks and Novel Dark Energy Relations}
Atomic clocks have become one of the most powerful precision tools in near-Earth astronomy and fundamental physics in recent times. Advances in optical lattice clocks, ion clocks and clock networks have pushed frequency stability and accuracy to the level of $10^{-17}$-$10^{-18}$ per year \cite{ac1Marion:2002iw,ac2Barontini:2021mvu,ac3Sherrill:2023zah,ac4Flambaum:2007my,ac5godun2014frequency}, and this has lead to enabling unique tests of possible variations in fundamental constants and of possible couplings between new fields and the Standard Model. In the context of modified gravity, atomic clocks have been used to constrain fifth forces, variations in particle masses, scalar couplings to electromagnetism, and violations of the Equivalence Principle \cite{ac6Uzan:2024ded,ac7Brzeminski:2022sde,ac8Levy:2024vyd,nato1Dent:2008vd,nato2Sherrill:2023zah,nato3Levy:2024vyd,nato4Elder:2025tue}. When interpreted in our context of a scalar-tensor dark energy framework, clock comparisons can provide a direct probe of the local time evolution of the scalar field responsible for cosmic acceleration and this in turn gives Solar System level constraints on the dark energy equation of state in combination with the LLR bounds derived previously. In this section we derive the relevant relations from first principles and obtain the equations that link clock measurements to the quantities $\dot\phi_0$, $w_0$, and $H_0$.
\\
\\
To establish the connection between clocks and scalar dynamics, we can consider that the scalar field generically induces a dependence of fundamental constants on $\phi$. For example, the electron mass $m_e$, the fine-structure constant $\alpha_{\rm EM}$ or the QCD scale may depend on $\phi$ through dimensionless couplings. Any atomic transition frequency $\nu_i$ in clock $i$ therefore inherits a dependence on the scalar field and so for small excursions of $\phi$ around its present value $\phi_0$, one may write
\begin{equation}
\nu_i(\phi) = \nu_i^{(0)}\left[1 + k_i\frac{\phi - \phi_0}{M_{\rm Pl}} + \cdots\right]
\end{equation}
where $k_i$ is the dimensionless sensitivity coefficient characterizing how strongly the transition in clock $i$ responds to changes in the scalar field and then taking the time derivative gives the fractional frequency drift,
\begin{equation}
\frac{\dot\nu_i}{\nu_i} \simeq k_i\,\frac{\dot\phi_0}{M_{\rm Pl}}
\end{equation}
One should note here that in practice, one never measures $\dot\nu_i$ for a single clock and instead, two clocks with different sensitivities $k_1$ and $k_2$ are compared after which we can define the observable quantity
\begin{equation}
D \equiv \frac{d}{dt}\ln\frac{\nu_2}{\nu_1}
= \frac{\dot\nu_2}{\nu_2} - \frac{\dot\nu_1}{\nu_1}
\end{equation}
Using this, one obtains the relation
\begin{equation}
D \simeq (k_2 - k_1)\frac{\dot\phi_0}{M_{\rm Pl}}
\label{eq:Dscalar}
\end{equation}
Letting $D_{\rm obs}$ denote the experimentally measured bound on this drift, one immediately gets
\begin{equation}
\dot\phi_0 = \frac{D_{\rm obs}}{k_2 - k_1}M_{\rm Pl}
\label{eq:phidot_clock}
\end{equation}
Current optical clock experiments place limits on $|D_{\rm obs}|$ at the level of $10^{-17}$-$10^{-18}\,{\rm yr}^{-1}$ and continued advances are expected to improve these bounds further in the not-so-distant future \cite{ac9Arvanitaki:2014faa,ac10Kennedy:2020bac,ac11Luo:2011cf}.
\\
\\
To relate these measurements to dark energy parameters, we use the cosmological relation for scalar kinetic energy obtained earlier \eqref{eq:phidot_background}. One can substitute the expression in \eqref{eq:phidot_clock} into \eqref{eq:phidot_background} which gives us
\begin{equation}
\left(\frac{D_{\rm obs}}{k_2 - k_1}M_{\rm Pl}\right)^2
= 3\,\Omega_{\rm DE,0}\,(1+w_0)\,M_{\rm Pl}^2 H_0^2
\end{equation}
Canceling $M_{\rm Pl}^2$ gives us the central clock constraint
\begin{equation}
H_0^2(1+w_0) =
\frac{D_{\rm obs}^2}{3\,\Omega_{\rm DE,0}\,(k_2-k_1)^2}
\label{eq:clock_constraint}
\end{equation}
This relation shows that atomic clocks do not independently determine $H_0$ or $w_0$ but what they do instead is they measure the single combination $H_0\sqrt{1+w_0}$
\begin{equation}
H_0\sqrt{1+w_0} =
\frac{|D_{\rm obs}|}{\sqrt{3\,\Omega_{\rm DE,0}}\;|k_2-k_1|}
\end{equation}
This shows that atomic clocks provide a direct local probe of the scalar kinetic energy today, they can allow one to separate $H_0$ and $w_0$ algebraically. Squaring both sides then allows us to write
\begin{equation}\label{eq:combined}
1+w_0 =
\frac{D_{\rm obs}^2}{3\,\Omega_{\rm DE,0}\,(k_2-k_1)^2}
\frac{1}{H_0^2}
\end{equation}
This formula represents the central result of our analysis and forms the basis of the constraints on scalar-tensor DE theories developed in the subsequent section.
\\
\\
\section{Constraints on Scalar-Tensor Dark Energy}
Now that we have established how one could use atomic clocks relations for considering the dark energy EOS, let us now figure out some constraints on dark energy models using the central relations derived earlier. As we shall see, the combination of the strong bounds on the EOS parameter from atomic clocks, alongside the solar system constraints on the running of the Planck mass, forces any scalar field that affects local gravitational physics to behave in an extremely restricted way today leading to powerful model-independent constraints on the structure of dark energy and the viability of many modified gravity scenarios.
\\
\\
We begin by considering the numerical implications of the combined relation \eqref{eq:combined}. To estimate the typical magnitude of the bound implied by \eqref{eq:combined}, we adopt indicative present day values directly entering the atomic clock relation. The Hubble constant lies in the interval $H_0\in[67,73]\ {\rm km\,s^{-1}\,Mpc^{-1}}$, corresponding to
\begin{equation}
H_0 \sim 2.2\times10^{-18}\ {\rm s}^{-1}
\end{equation}
For atomic clocks, a realistic differential drift constraint is
\begin{equation}
|D_{\rm obs}|\lesssim 10^{-17}\ {\rm yr}^{-1}\simeq 3\times10^{-25}\ {\rm s}^{-1}
\end{equation}
We further take $\Omega_{\rm DE,0}\simeq 0.69$ and an illustrative scalar sensitivity difference $|k_2-k_1|\sim 10^{-5}$, which is small enough to evade equivalence principle and fifth force constraints while remaining phenomenologically relevant. To be even more clear, this illustrative choice is physically motivated by atomic structure calculations which indicate that differential sensitivities of optical and hyperfine transitions to variations of fundamental constants typically lie in the range $10^{-6}$–$10^{-4}$ \cite{rosenband2008frequency}, depending on the species and transition involved. Taking $|k_2-k_1|\sim10^{-5}$ therefore represents a conservative intermediate value that is small enough to remain compatible with existing equivalence principle and fifth force bounds, while still capturing the level of sensitivity realistically accessible related future experiments. Substituting these values into the main relation \eqref{eq:combined}
one finds
\begin{equation}
1+w_0 \sim 10^{-4}
\end{equation}
with the precise value controlled by the clock sensitivity difference and the measured Hubble rate. Identifying a pair of clocks with a larger $|k_2-k_1|$ or adopting a slightly larger $H_0$ drives $1+w_0$ to even smaller values according to the scaling
\begin{equation}
1+w_0 \propto \frac{1}{(k_2-k_1)^2 H_0^2}
\end{equation}
In generic scalar-tensor theories admitting unsuppressed scalar couplings to atomic transitions, one therefore obtains a purely Solar System based bound
\begin{equation}
1+w_0 \lesssim 10^{-4}\text{-}10^{-5}
\end{equation}
which is significantly stronger than constraints derived from cosmological data alone, where simple late time parameterisations typically allow $|1+w_0|\lesssim 0.05$.
\\
\\
To understand the physical implications of such a tight bound, we recall that the scalar kinetic energy today is given by \eqref{eq:phidot_background}. Using $1+w_0\sim10^{-4}$, one finds
\begin{equation}
\frac{\dot\phi_0}{M_{\rm Pl}H_0} =
\sqrt{3\,\Omega_{\rm DE,0}\,(1+w_0)}
\sim 10^{-2}
\end{equation}
which implies that the scalar field evolves by only $\Delta\phi\sim \dot\phi_0/H_0\sim 10^{-2}M_{\rm Pl}$ over a Hubble time. Note again that this depends crucially on the estimate of the order of $(k_2-k_1)$ which we have taken over here due to considerations based on other solar system level probes and variation in fundamental constants. This corresponds not merely to slow roll but to an ultra slow roll regime, requiring the dark energy potential to be extremely shallow at the present epoch. For example, in a canonical quintessence model with an exponential potential of the form $V\propto e^{-\lambda\phi/M_{\rm Pl}}$, one has $1+w_0\simeq \lambda^2/3$, implying $\lambda\sim 10^{-2}$, which is far below the characteristic slopes typically invoked in quintessence constructions in the existing literature
\\
\\
The LLR bound $\alpha_{M,0}\lesssim 10^{-2}$, which we again note is consistent with PPN bounds too, also indicates that the effective Planck mass varies by at most order $1\%$ per Hubble time. If we want to talk about it in terms of the effective field theory of DE language then we say that this suppresses any modification of gravitational wave propagation, effective gravitational coupling, or gravitational slip which means that any scalar-tensor DE model with non-zero $\alpha_{M}$ must generate cosmic acceleration in a manner which should be nearly indistinguishable from a cosmological constant at the present epoch.
\\
\\
These combined constraints lead to significantly impacting the model space of dark energy as well. Consider Horndeski theories for example, which are the most general scalar-tensor theories with second-order equations of motion. Such theories generically predict non-zero $\alpha_M$ and field evolution sufficient to generate observable modifications to growth, lensing, or gravitational wave propagation. However, interesting late-time phenomenology frequently requires $|\alpha_{M,0}|\sim 0.05$-$0.2$ \cite{hd1Kase:2018aps,hd2Noller:2018wyv} and this is incompatible with the LLR bound $|\alpha_{M,0}|\lesssim10^{-2}$. Additionally, the scalar field motion required for nontrivial deviations from $w_0=-1$ is incompatible with the ultra slow roll bound implied by clocks and as a result, Horndeski dark energy must collapse into a regime essentially indistinguishable from $\Lambda$. This means that the LLR and atomic clock constraints together rule them out as a viable dark energy scenario as they are now. 
\\
\\
A similar argument applies to $f(R)$ gravity as well which is dynamically equivalent to a scalar-tensor theory with a Brans-Dicke scalar \cite{fr1Amendola:2006kh,fr2Battye:2015hza}. The combination of LLR and clock bounds requires the scalar degree of freedom $f_R$ to be nearly constant today, which means it is effectively forcing $f(R)$ to reduce to $R-2\Lambda$ at low redshift. Non-minimally coupled quintessence models where $F(\phi)R$ induces effective Planck mass evolution, also require substantial field motion to produce distinguishable late-time cosmology, yet the combined constraints force both $\dot\phi_0$ and $\alpha_{M,0}$ to be extremely small, eliminating these possibilities as well \cite{nc1Pace:2013pea,nc2Bhattacharya:2015wlz}. Brans–Dicke dark energy is similarly restricted as the Brans-Dicke scalar necessarily induces variations in $G$ that violate the LLR bound unless the Brans-Dicke parameter satisfies $\omega_{\rm BD}\gtrsim 10^4$ \cite{bd1Zucca:2019ohv}. In this limit the theory approaches general relativity and cannot support a dynamically evolving dark energy EOS distinct from $w_0=-1$.
\\
\\
Models invoking phantom dark energy driven by modified gravity often rely on substantial field motion at $z\simeq 0$ to generate significant deviations from $w_0=-1$ and the combined clock and LLR constraints eliminate these models as well \cite{pmg1Briscese:2006xu,pmg2Saratov:2012ni}, as they require $\dot\phi_0/(M_{\rm Pl}H_0)\sim{\cal O}(1)$ which is incompatible with the slow rolling  bound $\dot\phi_0/(M_{\rm Pl}H_0)\sim10^{-2}$. Finally, varying-$\alpha$ models in which the scalar field couples directly to electromagnetism, generically predict a non-zero drift of fundamental constants \cite{al2daFonseca:2022qdf}. Clock constraints require such drifts to be at most $10^{-17}$-$10^{-18}\ {\rm yr}^{-1}$ and combining this with the bound on $\dot\phi_0$ forces the scalar’s electromagnetic coupling to be effectively zero. This means that varying $\alpha$ dark energy models that do not rely on extreme screening are ruled out.
\\
\\
What these constraints mean is that any scalar field that modifies gravity or couples to matter strongly enough to be detectable locally must be very slowly rolling with negligible impact on gravitational dynamics today. In this framework, the surviving dark energy candidates are either a pure cosmological constant or minimally coupled quintessence models with negligible coupling to matter and no modification of the Planck mass. Scalar-tensor or modified gravity models that attempt to generate late time acceleration via field evolution or Planck mass running are severely constrained, and many widely studied classes of theories become incompatible with Solar System physics.
\\
\\
\section{Constraints on Non-Canonical Dark Energy}
Having studied the constraints on simple canonical scalar field theories, it is imperative for us to investigate beyond simple canonical theories too. Non-canonical scalar field theories of dark energy are characterized by a Lagrangian of the form $P(X,\phi)$ with $X\equiv-\frac{1}{2}g^{\mu\nu}\partial_\mu\phi\partial_\nu\phi$ and have been widely studied as alternatives to canonical quintessence. Such theories include k-essence, DBI-like models, ghost condensates, kinetic gravity braiding constructions and other extensions containing richer kinetic structure than the canonical $X - V(\phi)$ form. They are attractive because they can produce a wide range of cosmological behaviors, including variable sound speeds, non trivial EOS trajectories, NEC violating phases and in some cases they can address observational tensions. Because these models introduce new kinetic degrees of freedom, it is important to understand how strongly they are restricted by the combined Solar System constraints derived from LLR and atomic clock experiments. In this section we generalize our earlier canonical analysis to a fully non-canonical $P(X,\phi)$ setup, derive the corresponding relations for $w_0$ and $\dot\phi_0$ and discuss the consequences for several major classes of non-canonical dark energy.
\\
\\
We begin by taking a general scalar field Lagrangian
\begin{equation}
{\cal L}_\phi = P(X,\phi), \qquad
X = \frac{1}{2}\dot\phi^2
\end{equation}
on a homogeneous FLRW background again and the pressure and energy density are given by
\begin{equation}
p_\phi = P(X,\phi), \qquad
\rho_\phi = 2X P_X - P
\end{equation}
where $P_X\equiv \partial P/\partial X$ and so the DE EOS becomes
\begin{equation}
w_\phi = \frac{P}{2X P_X - P}, \qquad
1+w_\phi = \frac{2X P_X}{2X P_X - P}
\end{equation}
Evaluated today and identifying $\rho_{\rm DE,0}\simeq 3\Omega_{\rm DE,0}M_{\rm Pl}^2H_0^2$, one obtains
\begin{equation}
1+w_0 = \frac{2X_0 P_X(X_0,\phi_0)}{3\Omega_{\rm DE,0}M_{\rm Pl}^2H_0^2}
\label{eq:noncanon_w0_basic}
\end{equation}
with
\begin{equation}
X_0 = \frac{\dot\phi_0^2}{2}
\end{equation}
This generalizes the canonical relation $\dot\phi_0^2=3\Omega_{\rm DE,0}(1+w_0)M_{\rm Pl}^2H_0^2$, which is recovered when $P_X=1$.
\\
\\
We now use the atomic clock relation derived earlier and note that since the dependence of fundamental constants on $\phi$ is typically mediated directly by the field value rather than by its kinetic structure, the clock-side derivation remains unchanged even in the non-canonical case. One therefore has
\begin{equation}
D_{\rm obs} \equiv \frac{d}{dt}\ln\frac{\nu_2}{\nu_1} = (k_2-k_1)\frac{\dot\phi_0}{M_{\rm Pl}}
\end{equation}
which immediately implies
\begin{equation}
\dot\phi_0 = \frac{D_{\rm obs}}{k_2-k_1}M_{\rm Pl}, \qquad
X_0 = \frac{1}{2}\left(\frac{D_{\rm obs}}{k_2-k_1}\right)^2 M_{\rm Pl}^2
\label{eq:X0_clock_general}
\end{equation}
Substituting \eqref{eq:X0_clock_general} into \eqref{eq:noncanon_w0_basic} yields
\begin{equation}
1+w_0 =
\frac{D_{\rm obs}^2}{3\Omega_{\rm DE,0}(k_2-k_1)^2}\frac{P_X(X_0,\phi_0)}{H_0^2}
\label{eq:noncanon_clock_w}
\end{equation}
which shows explicitly that, in non-canonical theories, the deviation of the equation of state from $-1$ is controlled by the atomic clock sensitivity difference, the Hubble scale, and the kinetic prefactor $P_X$ evaluated on the present background. The dependence on $H_0$ is purely kinematical and does not rely on any assumption about the detailed functional form of $P(X,\phi)$.
\\
\\
We now evaluate the implications of \eqref{eq:noncanon_clock_w} using representative present-day values $H_0\simeq 2.2\times10^{-18}\,{\rm s}^{-1}$, $|D_{\rm obs}|\sim 3\times10^{-25}\,{\rm s}^{-1}$, $\Omega_{\rm DE,0}=0.69$, and an illustrative scalar sensitivity difference $|k_2-k_1|\sim 10^{-5}$. Inserting these values gives
\begin{equation} \label{noncan}
1+w_0 \sim 10^{-4}P_X(X_0,\phi_0)
\end{equation}
where all model dependence is encapsulated in the single multiplicative factor $P_X(X_0,\phi_0)$ evaluated on the present background. The canonical limit is smoothly recovered for $P_X=1$, while deviations from unity directly rescale both the magnitude of $1+w_0$ and the tightness of the Solar System bound in non-canonical dark energy scenarios
\\
\\
This tells us that if $P_X$ is of order unity, the bound on $1+w_0$ remains at the $10^{-4}$-$10^{-5}$ level. If $P_X\gg 1$, which is true in some DBI or strongly non-canonical k-essence models, then the allowed parameter space contracts even further as either $(k_2-k_1)$ or $\alpha_{M,0}$ must be suppressed to maintain the same bound on $1+w_0$. Conversely in ghost condensate type theories where $P_X\approx 0$ at a special point, the EOS naturally approaches $w_0\simeq -1$, but this requires the field to sit exactly at the condensate point and remain extremely stable and so any drift away from that point would be detected by clocks through $\dot\phi_0$.
\\
\\
We can now talk in general about the consequences here for several broad classes of non-canonical dark energy. Dynamical k-essence models typically require $P_X\sim{\cal O}(1)$ \cite{ke1Chimento:2004jm,ke2Chakraborty:2019swx,ke3Li:2006bx}and field velocities $\dot\phi_0\sim M_{\rm Pl}H_0$ to generate significant deviations from $w_0=-1$, but the combined LLR and clock bounds enforce $\dot\phi_0\sim 10^{-2}M_{\rm Pl}H_0$ and $1+w_0\lesssim 10^{-4}$, which ends up ruling out such behavior. DBI dark energy which involves a relativistic kinetic structure with $P_X=\gamma\ge 1$, faces an even sharper tension as large $P_X$ amplifies its predicted $(1+w_0)$ for a given field velocity \cite{dbi1Kaeonikhom:2012xr,dbi2Fahimi:2018pcr}, making it incompatible with the small Solar System allowed values unless couplings to matter and gravity are tuned to zero. Ghost condensate theories survive only in the special case where the field is exactly at the condensate point today so that $P_X\approx 0$ and $1+w_0\simeq 0$ automatically \cite{gc1Piazza:2004df,gc2Bhattacharya:2013vv,gc3Ashoorioon:2023zju}. Any nonzero motion would be detected by clocks and ruled out unless the field is perfectly stabilized.
\\
\\
Tachyonic models of dark energy which rely on non-standard kinetic structure to achieve accelerated expansion, generally require substantial scalar motion or large kinetic coefficients and these conditions violate the LLR and clock bounds on $\dot\phi_0$ and $1+w_0$ \cite{tachBagla:2002yn}, leaving only the trivial case unscathed where the field behaves effectively as a cosmological constant. Phantom k-essence models which require large deviations from $w=-1$ or even $w<-1$, are excluded for the same reason as well as the combined constraints force $|1+w_0|\lesssim 10^{-4}$, making phantom behavior impossible unless completely decoupled from local gravity. Kinetic gravity braiding models \cite{kgb1Deffayet:2010qz,kgb2BorislavovVasilev:2024loq} which rely on a non-zero braiding parameter and mixing between $\phi$ and metric derivatives, predict nontrivial late-time dynamics of the effective Planck mass and so the LLR bound $|\alpha_{M,0}|\lesssim 10^{-2}$ rules out the required regime for these models to have observable modified gravity effects.
\\
\\
The main insight we get here is that if the same scalar affects both the dark energy background and local gravitational or matter couplings, then LLR and atomic clock constraints enforce a very slowly rolling  regime with $1+w_0\lesssim 10^{-4}$-$10^{-5}$, irrespective of the kinetic structure. This rules out dynamical k-essence, DBI, tachyonic, phantom k-essence and kinetic gravity braiding dark energy models in this framework. Only two non-canonical possibilities remain viable, with those being that of k-essence with negligible local coupling and ghost condensates exactly at their condensate point, but it is easy to see that these are very special cases and one could even say ad-hoc. Otherwise, all non-canonical scalar-tensor DE models collapse into a regime effectively indistinguishable from a cosmological constant at the present epoch.
\\
\\
\section{Conclusions}
It is important here to clarify and discuss one subtlety in our work firstly. A key assumption underlying our analysis is that the same scalar field responsible for cosmic acceleration may also modify gravity locally and couple to matter through variations in fundamental constants and this notion is well motivated across a wide class of dark energy constructions. In scalar–tensor theories, we see that non-minimal couplings $F(\phi)R$ naturally lead to a time dependent effective Planck mass implying that the scalar field driving cosmic expansion inevitably leaves imprints on local gravitational dynamics unless screened. Likewise one can see that in many high energy embeddings like dilaton models, moduli fields, quintessence coupled through effective operators etc. light scalar fields generically mediate variations in particle masses or gauge couplings. These interactions cannot be independently tuned away without eliminating the mechanisms that allow the scalar to influence cosmic expansion. This means that the assumption that the dark energy scalar simultaneously affects cosmic acceleration, the local value of Newton’s constant, and the physics underlying atomic transitions is not only natural but representative of broad and theoretically motivated sectors of modified gravity and scalar DE theories. Precisely because this assumption holds across such a large portion of the theoretical landscape, Solar System probes like LLR and high precision atomic clocks become powerful tools for constraining DE physics at $z\simeq 0$.
\\
\\
To conclude, in this work we have shown that atomic clock experiments provide a powerful and conceptually clean local probe of late-time dark energy dynamics in scalar–tensor theories. Starting from first principles, we derived a direct mapping between differential clock frequency drifts and the present day scalar field velocity $\dot\phi_0$, allowing atomic clock observables to be translated into constraints on the kinetic energy of the dark energy sector and we found master relation of the form
\begin{equation*}
    1+w_0 =
\frac{D_{\rm obs}^2}{3\Omega_{\rm DE,0}(k_2-k_1)^2}
\frac{1}{H_0^2}
\end{equation*}
for canonical models, with a straightforward extension involving the kinetic prefactor $P_X$ in non-canonical theories. Existing bounds from Lunar Laser Ranging on the running of the effective Planck mass provide an important consistency check on the allowed scalar dynamics and help anchor the local gravitational sector, but the dominant constraining power in our analysis arises from atomic clock measurements themselves.
\\
\\
For realistic clock sensitivities and physically motivated scalar–matter couplings, we find that atomic clocks alone already force the present-day dark energy equation of state into the narrow range $1+w_0\lesssim10^{-4}$–$10^{-5}$ and this corresponds to an ultra-slow-roll regime with $\dot\phi_0/(M_{\rm Pl}H_0)\sim10^{-2}$ or smaller, implying that any scalar field responsible for cosmic acceleration must evolve extremely slowly at $z\simeq0$. Extending the analysis to non-canonical dark energy models shows that this conclusion is strong up to multiplicative factors involving $P_X$, which means that a wide class of theories including k-essence, DBI-type models, ghost condensates tachyonic constructions and kinetic gravity braiding are quite tightly constrained by precision atomic clock experiments, with only highly tuned scenarios remaining viable.
\\
\\
A valuable point to emphasize is that there are currently no conclusive, direct, model independent experimental constraints on the scalar sensitivity difference $(k_2-k_1)$ itself, and the value adopted in this work should be understood as an illustrative order of magnitude estimate rather than a measured quantity. The choice $|k_2-k_1|\sim10^{-5}$ is motivated by being sufficiently small to remain compatible with existing equivalence principle and fifth-force bounds, while still being large enough to allow atomic clock experiments to probe cosmologically relevant scalar dynamics and in this sense, the resulting bounds on $1+w_0$ should be interpreted not as definitive exclusions, but as demonstrating the potential constraining power of atomic clocks on dark energy once $(k_2-k_1)$ is better determined. Future improvements in precision combined with advances in atomic structure calculations and searches for variations of fundamental constants, will allow $(k_2-k_1)$ to be constrained more robustly and will thereby help in converting the relations derived here into more direct and competitive probes of late time dark energy dynamics. In particular, ongoing and forthcoming efforts involving next generation optical clock networks \cite{rosenband2008frequency,Hees:2016gop}, space based clock missions and large scale surveys sensitive \cite{Herrmann:2018rva,Uzan:2010pm} to variations of fundamental constants are expected to play a central role in sharpening these constraints.
\\
\\
These results collectively point toward a strong and general conclusion being that any scalar field that both drives cosmic acceleration and interacts with local gravitational or matter sectors must behave almost exactly like a cosmological constant at the present epoch. This significantly narrows the viable phase space of scalar tensor dark energy and eliminates most late time modified gravity explanations of phenomena such as the Hubble tension or deviations in structure growth. The framework developed here invites several future directions as well like including extending these constraints using next generation optical clocks, space based clock networks etc. This also provides a ruling that one interested in addressing cosmic tensions of the current age like $H_0$ tension by considering non-trivial dark energy models, should make sure to make models which are also consistent from Solar System level constraints.

\section*{Acknowledgments}
The author would like to thank Robert Scherrer for very helpful discussions on this work. The author would also like to thank the Discovery Doctoral Fellowship at Vanderbilt for supporting the work.

\bibliography{references}
\bibliographystyle{unsrt}

\end{document}